# Structural and superconducting properties of RbOs$_2$O$_6$ single crystals

Krzysztof Rogacki[1,2,*], Götz Schuck[3], Zbigniew Bukowski[1], Nikolai D. Zhigadlo[1], and Janusz Karpinski[1]

[1] *Laboratory for Solid State Physics, ETH Zürich, 8093 Zürich, Switzerland*
[2] *Institute of Low Temperature and Structure Research, PAS, Wroclaw, Poland*
[3] *Laboratory for Neutron Scattering, PSI, 5232 Villigen, Switzerland*

## Abstract

Single crystals of RbOs$_2$O$_6$ have been grown from Rb$_2$O and Os in sealed quartz ampoules. The crystal structure has been identified at room temperature as cubic with the lattice constant $a$ = 10.1242(12) Å. The anisotropy of the tetrahedral and octahedral networks is lower and the displacement parameters of alkali metal atoms are smaller than for KOs$_2$O$_6$, so the "rattling" of the alkali atoms in RbOs$_2$O$_6$ is less pronounced. Superconducting properties of RbOs$_2$O$_6$ in the mixed state have been well described within the London approach and the Ginzburg-Landau parameter $\kappa(0) \cong 31$ has been derived from the reversible magnetization. This parameter is field dependent and changes at low temperatures from $\kappa \cong 22$ (low fields) to $\kappa \cong 31$ at $H_{c2}$. The thermodynamic critical field $H_c(0)$ = 1.3 kOe and the superconducting gap $2\Delta_o/k_BT_c \cong 3.2$ have been estimated. These results together with slightly different $H_{c2}(T)$ dependence obtained for crystals and polycrystalline RbOs$_2$O$_6$ proof evidently that this compound is a weak-coupling BCS-type superconductor close to the dirty limit.

*Corresponding author:
K. Rogacki, Rogacki@int.pan.wroc.pl

I. INTRODUCTION

The recent discovery of superconductivity in a $\beta$-pyrochlore compound $KOs_2O_6$ (Ref. 1) has renewed both theoretical and experimental interest in the properties of superconductors with geometrical frustration in a triangular lattice, which may give rise to many interesting phenomena including "rattling" behavior of atoms in spacious lattice sites.[2,3] There is strong experimental evidence that superconductivity in $KOs_2O_6$ is unconventional[4,5,6] and possibly related to its non-centrosymmetric structure[7]. Thus, unconventional pairing has been considered for this compound, which is superconducting at $T_c$ = 9.6 K and shows an extremely large upper critical field at zero temperature $H_{c2}(0) \approx 300$ kOe.[5,8] This field is much higher than the paramagnetic limiting field in the BCS approach, $H_p(0) = 18.4 T_c \approx 180$ kOe, indicative of pairing with the p-wave symmetry, however an explanation considering s-wave paring and extremely strong electron-phonon coupling has been proposed as well.[4,6] The list of superconducting $\beta$-pyrochlores includes also $CsOs_2O_6$ and $RbOs_2O_6$ with $T_c \approx 3$ and 6 K, respectively. For these compounds, which until now have been available in a polycrystalline form only, no inversion symmetry breaking has been found[9] and a rather conventional weak-coupling[10,11,12] or intermediate-coupling[13] BCS-type superconductivity has been reported. However, for $RbOs_2O_6$, more unusual mechanism of superconductivity has been also considered on the basis of specific heat measurements.[14,15] Since thermodynamic properties of the $\beta$-pyrochlore superconductors are sensitive to impurities and other inhomogeneities,[16] and since very often the polycrystalline $RbOs_2O_6$ contains an amount of trace impurities,[13,15,17,18,19] studies of high quality single crystals are important to avoid possible ambiguity for obtained results. Moreover, it was pointed out that for polycrystalline pyrochlore materials, some normal-state and superconducting properties can not be estimated properly because of an uncertainty from the grain-surface effects.[19]

The crystal structure of $\alpha$-pyrochlore $A_2B_2O_6O'$ is cubic, where atoms from each A and B sites constitute a corner sharing 3D tetrahedral network called the pyrochlore network. For the $\alpha$-pyrochlore with $Fd\bar{3}m$ symmetry, all tetrahedra of each sublattice are crystallographic equivalent. This is not the case for the low temperature phase of superconducting $Cd_2Re_2O_7$, were two different Re and Cd tetrahedra are present (breathing mode like behavior), so the breaking of the inversion symmetry is observed.[20] The $\beta$-pyrochlore $\square_2B_2O_6A'$ (B = Os; A' = K, Rb) structure can be described as a modified $\alpha$-pyrochlore structure were β-pyrochlore A' atoms occupy the O' sites and the $\alpha$-pyrochlore A sites are empty. Recently, the crystal structure of $RbOs_2O_6$ and $KOs_2O_6$ was described within the $Fd\bar{3}m$ symmetry.[9,21] However, for high quality $KOs_2O_6$ single crystals, the non-centrosymmetric $F\bar{4}3m$ space group was also proposed in order to explain the additional Bragg reflections, which violated the $Fd\bar{3}m$ symmetry.[7]

In this paper we examine the structural and superconducting properties of high quality $RbOs_2O_6$ single crystals. We show that the centrosymmetric $\beta$-pyrochlore structure is violated at room temperature, however the atomic shifts from the centrosymmetric positions are lower than those reported for $KOs_2O_6$. The superconducting properties of $RbOs_2O_6$ crystals are rather conventional and differ much from those observed for $KOs_2O_6$. We have considered the possible reason for the observed differences and found that the "rattling" intensity of the alkali metal atoms seems to be the most important parameter which controls the superconducting properties of the $\beta$-pyrochlores. Such a statement is recently beginning to emerge from both experimental and theoretical studies, as implied in Ref. 3, 6, 19, and 22.

## II. EXPERIMENTAL

Single crystals of $RbOs_2O_6$ have been grown from stoichiometric amounts of Os metal (Alfa Aesar, 99.9%) and $Rb_2O$ (Aldrich). The components were thoroughly mixed and pressed into pellets in a dry box and sealed in an evacuated quartz ampoule together with an appropriate amount of $Ag_2O$ (Aldrich, 99%) as a source of oxygen. The ampoule was placed in a preheated furnace at 600°C, kept for 10 -15 h then slowly (5 °C/h) cooled to 400°C and finally fast cooled to room temperature. Because of strong reaction of the pellets with quartz, alumina crucibles have been used to isolate the pellets from the ampoule walls. Single crystals of $RbOs_2O_6$ were found on the walls of the alumina crucible indicating that the crystals grew from a gas phase. The crystals have shapes of flattened octahedra and sizes up to 0.3 x 0.4 x 0.4 $mm^3$ (Fig. 1). Energy dispersive X-ray analysis (EDX) indicate slight Rb deficiency but this result should be taken with caution because of strong overlapping of Os and Rb lines in the EDX spectrum.

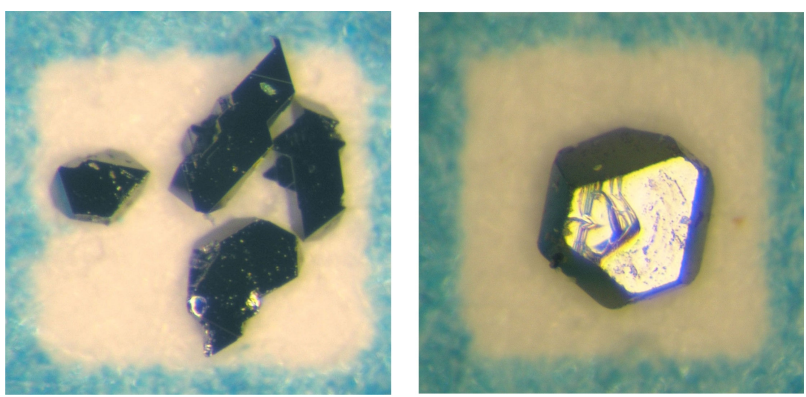

FIG. 1. (Color online) Single crystals of $RbOs_2O_6$ grown in quartz ampoules. The edge of each individual picture is about 1 mm.

Structural properties of the $RbOs_2O_6$ single crystals have been investigated at room temperature with an X-ray single crystal diffractometer [Siemens SMART charge coupled device (CCD) system] using graphite-monochromated Mo $K\alpha$ radiation.[23] 1818 frames were measured with an exposure time of 10 s per frame and $\Delta\Omega = 0.3°$. The cubic unit cell parameter $a = 10.1242(12)$ Å was refined from 4094 reflections. The three-dimensional data were reduced and corrected for Lorentz, polarization and background effects using SAINT and for absorption and the $\lambda/2$ contamination using SADABS.[23,24] In order to verify the correction for the $\lambda/2$ contamination, a similar procedure was applied for an $MgB_2$ single-crystal dataset obtained with the same diffractometer and the results have confirmed the reliability of our correction. The structure was solved by direct methods using SHELXS-97.[25] The fullmatrix least-squares refinement on $F^2$ was achieved with SHELXL-97.[25,26]

Superconducting properties of the $RbOs_2O_6$ single crystals have been investigated by measurements of the magnetic moment, $m$, as a function of temperature and magnetic field employing a Quantum Design SQUID magnetometer (MPMS-5) equipped with a 5 T magnet. Individual crystals with a mass of about 40 µg as well as a set of 2 crystals with a mass of 82 µg were studied to obtain quantitative results. The magnetic moment $m$ was measured at constant field upon heating from zero-field-cooled state (ZFC mode) and upon cooling (FC mode), or at constant temperature with increasing and decreasing field. The temperature sweep was about 0.2 K/min resulting in an accuracy with which $m$ is measured better than $1 \cdot 10^{-6}$ emu, also in high magnetic fields, for an option where results of 3 scans were averaged to obtain each individual experimental point.

III. CRYSTAL STRUCTURE

Low-intensity X-ray reflections forbidden in the $\beta$-pyrochlore structure with the centrosymmetric space group $Fd\bar{3}m$ have been observed for our $RbOs_2O_6$ crystals at room temperature. These reflections were present despite the fact that the correction for the $\lambda/2$ contamination has been performed. After averaging symmetry-equivalent reflections, nine reflections inconsistent with d-glide symmetry (0$kl$: $k+l \neq 4n$, e.g., 024) and four reflections inconsistent with $4_1$ screw axis (00$l$: $l \neq 4n$, e.g., 002) of $Fd\bar{3}m$ persist [$I > 3\sigma(I)$]. This clearly indicates that the previously proposed symmetry of the $Fd\bar{3}m$ space group is broken. No indication for a tetragonal splitting and no reflections which break the F-centring ($h,k,l$ not all even or not all odd) have been found. The observed symmetry lowering results in the cubic non-centrosymmetric space group $F\bar{4}3m$. The Os atom has been located at the position 16$e$ using the direct method. Positions of other atoms were derived from results obtained for $KOs_2O_6$.[7] The high quality of $RbOs_2O_6$ crystals allowed us to refine anisotropic displacement parameters (ADP) for all atoms. As a result, the final refinement yielded a residual of $R_1 = 0.0194$. Results of the X-ray measurements and refined atomic parameters are given in Table I and Table II. Selected interatomic distances and bond angles are given in Table III. Some structural details are shown in Fig. 2.

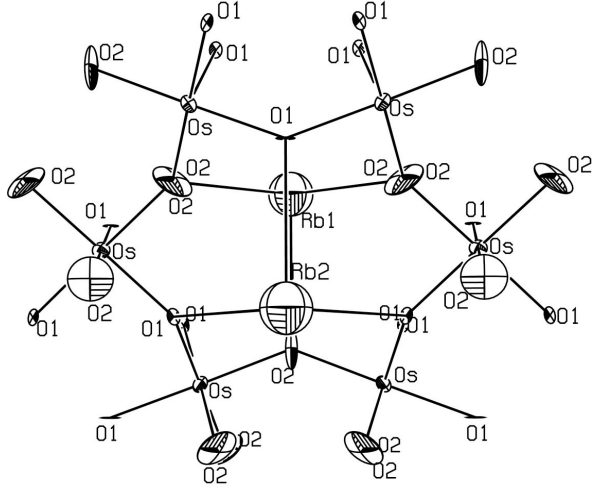

FIG. 2. Structural details of RbOs$_2$O$_6$ at room temperature ($F\bar{4}3m$ space group) showing the anisotropic nature of the rubidium channels (structure projected along the [-110] direction). Interatomic distances Rb1-O2 and Rb2-O1 are shorter (3.1081 Å) and longer (3.1395 Å), respectively, compared to the $\beta$-pyrochlore structure with $Fd\bar{3}m$ symmetry [Rb-O = 3.125(7) Å]. The O2 displacement ellipsoid is elongated toward the Rb1 atom whereas the O1 displacement ellipsoid does not point to the Rb2 atom. The displacement ellipsoids are drawn at the 50% probability level.

The Rb1-O2 bond length is shorter than the Rb2-O1 bond length (the Rb-O bond length difference $\Delta_{Rb-O}$ = 0.0314 Å) and, consequently, the Os-O2 bond length is longer than Os-O1 ($\Delta_{Os-O}$ = 0.026 Å). This leads to anisotropic alkali metal channels (Fig. 2 and Table III), where Rb2 atoms have more space to move, so they oscillate with a larger amplitude. The anisotropy of the alkali metal channels is also reflected in the refined ADP`s of the Rb and O atoms. The ADP`s of Rb2 and O2 are much larger than the ADP`s of Rb1 and O1. Moreover the O2 displacement ellipsoid is elongated toward the Rb1 atom whereas the O1 displacement ellipsoid is not oriented in the consistently similar way, i.e., this ellipsoid does not point to the Rb2 atom. Summarizing, the lowering of symmetry from $Fd\bar{3}m$ to $F\bar{4}3m$ results in the bond length differences in both OsO$_6$ and O$_6$ octahedral networks ($\Delta_{Os-O}$ = 0.026 Å, $\Delta_{O-O}$ = 0.041 Å) leading to a breathing mode like behaviour. Refined ADP's, $\Delta$'s, and $U_{iso}$'s show that the anisotropy of the alkali metal channels and the "rattling" of the alkali metal atoms obtained for RbOs$_2$O$_6$ are much lower than those for KOs$_2$O$_6$ (see Table II and Table III). This may explain the exceptionally different superconducting properties of the KOs$_2$O$_6$ compound if compare to properties of the other members of the $\beta$-pyrochlore family.

As mentioned in the Introduction, the crystal structure of RbOs$_2$O$_6$ has been previously described by Galati et al.[9] within the centrosymmetric space group $Fd\bar{3}m$. This result has been obtained from the powder neutron diffraction (PND) experiments, where no Bragg reflections forbidden for the centrosymmetric structure have been observed.

Considering the resolution of the PND measurements, we may speculate that the low-intensity reflections, which violate the $Fd\bar{3}m$ symmetry, are hidden in the background. The background may be additionally enhanced for polycrystalline samples where some amount of trace impurities are usually present.[13,15,17-19] Consequently, the evidence for the inversion symmetry breaking in the β-pyrochlore compounds may be observed for very pure material only, particularly in the single-crystal form.

## IV. SUPERCONDUCTING PROPERTIES

The superconducting transition temperature $T_c$ was determined from changes of the magnetic moment $m$ with temperature at constant field. As an example, Figs. 3 and 4 show $m(T)$ results for two $RbOs_2O_6$ single crystals with similar $T_c \approx 6.2$ K. The transition temperature, $T_c$, and the transition onset temperature, $T_{co}$, were defined as illustrated in Fig. 3. The transition temperature difference, $\Delta T_c = T_{co} - T_c$, was determined and identified with crystal quality. This temperature difference varied from 0.05 to 0.15 K, depending on a batch. For high quality crystals, $\Delta T_c \approx 0.05$ K is roughly field independent at low fields and drops below the resolution of $\Delta T_c \approx 0.02$ K at $H \geq 4$ kOe. This seems to confirm a pure bulk transition (no surface effects) to the superconducting state at $T_c$. The presence of $T_{co}$ at low fields can be explain as a result of inhomogeneous field distribution around the crystal due to a large demagnetizing effect expected for the octahedral-shaped crystal. Thus $\Delta T_c$ should increase when the quality of the crystal surface lowers.

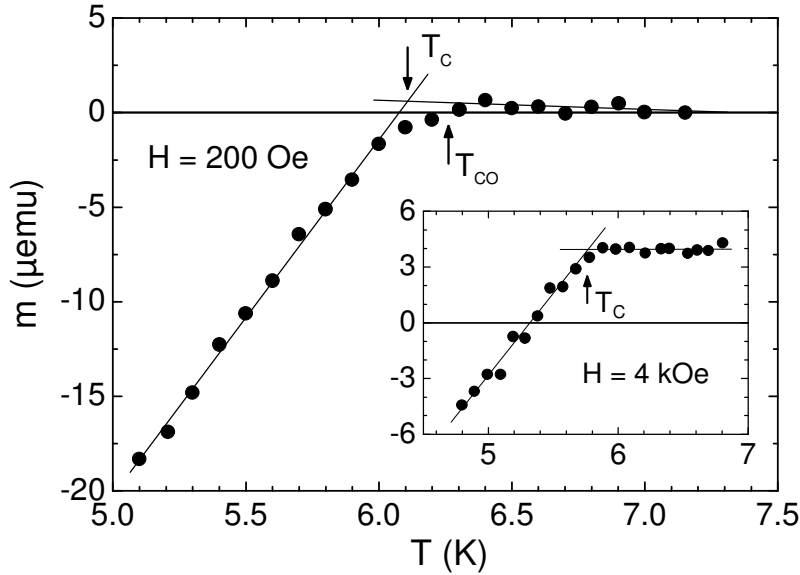

FIG. 3. Magnetic moment vs temperature at constant field $H = 0.2$ kOe and 4 kOe (inset) for the $RbOs_2O_6$ single crystal with $T_c = 6.25$ K at 5 Oe. The superconducting transition temperature, $T_c$, and the transition onset temperature, $T_{co}$, are marked by arrows. For $H = 4$ kOe, $T_{co} \approx T_c = 5.76$ K indicative of a bulk transition to the superconducting state.

In Fig. 4 we show the magnetic moment of the $RbOs_2O_6$ single crystal measured with increasing (ZFC mode) and decreasing (FC mode) temperature at a field of 500 Oe. A sharp transition to the superconducting state, observed at $T_c = 6.18$ K in 5 Oe (see inset), broadens significantly with increasing field and extends to the lowest available temperatures for $H \geq 200$ Oe. This indicates weak pinning properties and, consequently, a low irreversibility field, $H_{irr}$, even at low temperatures. For example, at a temperature of 3.4 K the upper critical field $H_{c2} = 28.5$ kOe, however $H_{irr}$ is equal only to about 2 kOe, for the critical current criterion $j_c \approx 6$ A/cm$^2$ which corresponds to the magnetic moment resolution $\Delta m \approx 5 \cdot 10^{-7}$ emu. The extremely low pinning, and thus $H_{irr}$, seems to be a more general feature of the $\beta$-pyrochlore compounds, as analogous weak pinning properties have been recently reported for $KOs_2O_6$ crystals.[7]

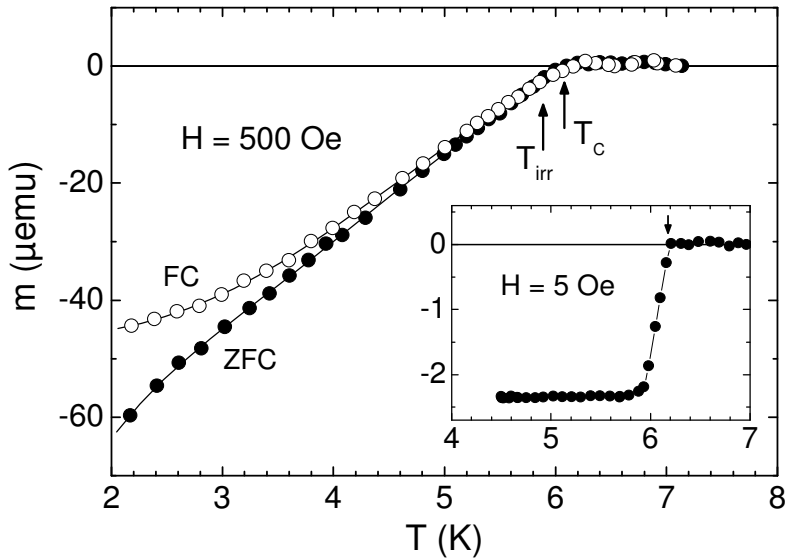

FIG. 4. Magnetic moment vs temperature at constant field $H = 500$ Oe and 5 Oe (inset) for the $RbOs_2O_6$ single crystal with $T_c = 6.18$ K at 5 Oe. The results are obtained upon heating from zero-field-cooled (ZFC) state and upon cooling (FC). The superconducting transition temperature, $T_c$, and the irreversibility temperature, $T_{irr}$, are marked by arrows. The beginning of the irreversible $m(T)$ properties at $T_{irr}$ is not visible in this scale.

Results of the magnetic moment measurements similar to those presented in Figs. 3 and 4 have been performed at constant fields up to 50 kOe to obtain the temperature dependence of the upper critical field $H_{c2}(T)$. In Fig. 5 we show the $H_{c2}$-$T$ phase diagram for the $RbOs_2O_6$ crystal with $T_c = 6.18$ K at 5 Oe (see the inset of Fig. 4). Special attention has been paid to determine $H_{c2}(T)$ in the vicinity of $T_c$ resulting in the real initial slope $(-dH_{c2}/dT)_{Tc} = 11.0$ kOe/K. This value is slightly lower than 12 kOe/K reported for polycrystalline samples.[18] The notable increase of the initial slope of $H_{c2}(T)$ observed for polycrystalline $RbOs_2O_6$ can be explained as a result of the enhanced scattering of quasiparticles due to structural imperfections. This may suggest that the RbOs$_2$O$_6$ superconductor is close to the dirty limit. In the inset of Fig. 5, some details of the $H_{c2}(T)$ dependence are shown near $T_c$. The linear extrapolation of $H_{c2}(T)$ to $H_{c2} = 0$ results in $T_c$ = 6.13 K, which is a slightly lower temperature than $T_c$ = 6.18 K obtained from the $m(T)$ measurements performed at $H$ = 5 Oe. The same $T_c$ = 6.18 K can be found by the linear extrapolation of the $H_{c2}(T)$ dependence obtained for the transition onset temperature $T_{co}$ (see the inset of Fig. 5). This shows that for our single crystals, the possible surface effects can notably influence the superconducting properties only at low fields. For fields higher than about 100 Oe, the bulk transition to the superconducting state is clearly marked at $T_c$.

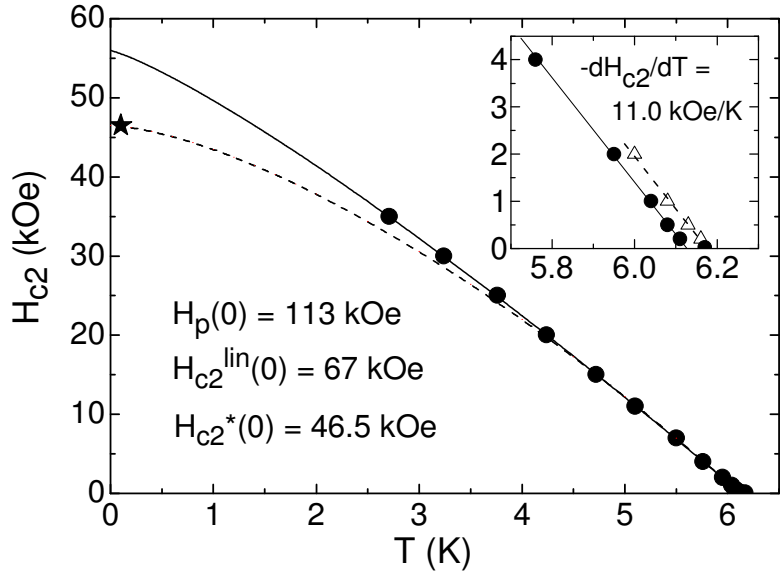

FIG. 5. Upper critical field, $H_{c2}$, extracted from the magnetization measurements (circles) for the RbOs$_2$O$_6$ single crystal with a bulk transition temperature $T_c$ = 6.13 K. The solid line is the power law fit $H_{c2}(T) = H_{c2}(0)[1-(T/T_c)^n]$ with an exponent $n$ = 1.2 and $H_{c2}(0)$ = 56 kOe. The dashed line is the fit with $n$ = 1.5 and $H_{c2}^*(0)$ = 46.5 kOe (star), where $H_{c2}^*(0)$ is the GLAG orbital limiting field in the dirty limit. The inset shows the initial $H_{c2}(T)$ dependence derived with the bulk (circles) and the onset (triangles) transition temperatures. The BCS Pauli limiting field, $H_p(0)$ = 113 kOe, is much higher than $H_{c2}^{lin}(0)$ = 67 kOe obtained from a linear extrapolation of the initial $H_{c2}(T)$ dependence.

The orbital upper critical field at zero temperature, $H_{c2}^*(0)$, can be obtained in the frame of the Ginzburg-Landau-Abrikosov-Gor'kov (GLAG) theory in the dirty limit from a formula: $H_{c2}^*(0) = -0.69(dH_{c2}/dT)_{Tc} \cdot T_c$.[27] Thus, $H_{c2}^*(0)$ = 46.5 kOe has been calculated for the RbOs$_2$O$_6$ crystals with $T_c$ = 6.13 K. This value has been obtained with an assumption that the spin-orbit scattering parameter $\lambda_{so} = \infty$, thus the superconducting pairs are broken due to the orbital effects only. The Pauli paramagnetic limiting field in the BCS Clogston approach is $H_p(0) = 18.4 \cdot T_c \cong 113$ kOe and, consequently, the Maki parameter $\alpha_M = \sqrt{2} \cdot H_{c2}^*(0)/H_p(0) \cong$ 0.58.[28,29] This suggests that the spin contribution to the pair-breaking effect is weak and can not influence $H_{c2}$ significantly.[27,29,30] In consequence, we assume that $H_{c2}(0) \approx H_{c2}*(0) = 46.5$ kOe and then the simple empirical formula $H_{c2}(T) = H_{c2}(0)[1-(T/T_c)^n]$ with $n = 1.5$ is used to obtain the best fitting to the experimental data, as shown in Fig. 5. However, for fields higher than 25 kOe, the calculated curve differs from the experimental data evidently. Similar deviation to higher values of the experimental $H_{c2}(T)$ from the GLAG formula has been recently reported for polycrystalline $RbOs_2O_6$.[18] For this material, extrapolated $H_{c2}(0) = 56$ kOe has been obtained from resistivity measurements performed down to 1 K.[16] For our single crystals, the best fitting of the empirical formula $H_{c2}(T) = H_{c2}(0)[1-(T/T_c)^n]$ to the experimental points has been found for $H_{c2}(0) = 56$ kOe and $n = 1.2$, exactly as for polycrystalline samples. Note, that even a linear extrapolation of $H_{c2}(T)$ from $T_c$, where -$dH_{c2}/dT = 11.0$ kOe/K, to $T = 0$ K results in $H_{c2}^{lin}(0) \cong 67$ kOe, which is a much lower value than $H_p(0) \cong 113$ kOe. Thus, we can conclude that $RbOs_2O_6$ is in the orbital limit, much away from the paramagnetic limit, and the extrapolated $H_{c2}(0) = 56$ kOe can be used to derive the GL coherence length at zero temperature $\xi(0) = [\Phi_0/2\pi H_{c2}(0)]^{1/2} \cong 7.7$ nm, where $\Phi_o$ is the magnetic flux quantum. Then, the GL parameter can be estimated $\kappa = \lambda(0)/\xi(0) \cong 32$, with the GL penetration depth $\lambda(0) \cong 250$ nm obtained in the recent muon-spin-rotation experiments for polycrystalline $RbOs_2O_6$.[10,11] This relatively large $\kappa$ enables us to evaluate the lower critical field by using the BCS formula $H_{c1}(0) = \Phi_0 \ln\kappa/4\pi\lambda^2 \cong 91$ Oe. This field is only slightly higher than the first penetration field, $H_{fp}$, derived from the $m(H)$ measurements, as we discuss in the next paragraph. Similar $H_{c1}(0) = 92$ Oe was obtained for the polycrystalline compound from the specific heat measurements.[13]

Magnetic moment versus field has been measured at constant temperature to examine the reversible magnetization, $M_{rev}$, the first penetration field, $H_{fp}$, and the shielding effect, simultaneously for two $RbOs_2O_6$ crystals with masses of 51 and 31 μg, and dimensions of roughly 0.12 x 0.2 x 0.3 and 0.12 x 0.12 x 0.3 mm³, respectively. Figure 6 shows the magnetization loop, $M(H)$, obtained at 3.4 K for $H$ swept between 1.5 kOe and -1.5 kOe, beginning from zero. This loop is very narrow and signifies weak pinning and low irreversibility field, consistently with the $m(T)$ results. For $T = 3.4$ K, this field is $H_{irr} \approx 2$ kOe, despite of much higher $H_{c2} = 28.5$ kOe. The beginning of the magnetization loop (virgin curve) is shown in the inset (a) of Fig. 6. First deviation from the linear part of the $M(H)$ curve is used to obtain $H_{fp}$, which is an upper limit of the lower critical field, $H_{c1}$. For $T = 3.4$ K, the first deviation is observed at $H = 41$ (±1) Oe (see the inset), and after correction for demagnetizing effects this value results in $H_{fp} = 82$ Oe. For correction we used a demagnetizing factor $N = 0.5$, for $H$ perpendicular to the [111] crystallographic direction of the octahedral-shaped crystals.[7] For this demagnetizing factor, and for the theoretical density of 7.15 g/cm³ for $RbOs_2O_6$, the superconducting volume fraction $V_s = 0.97$ (±0.03) has been estimated. This confirms full diamagnetism of the crystals certifying their good quality.

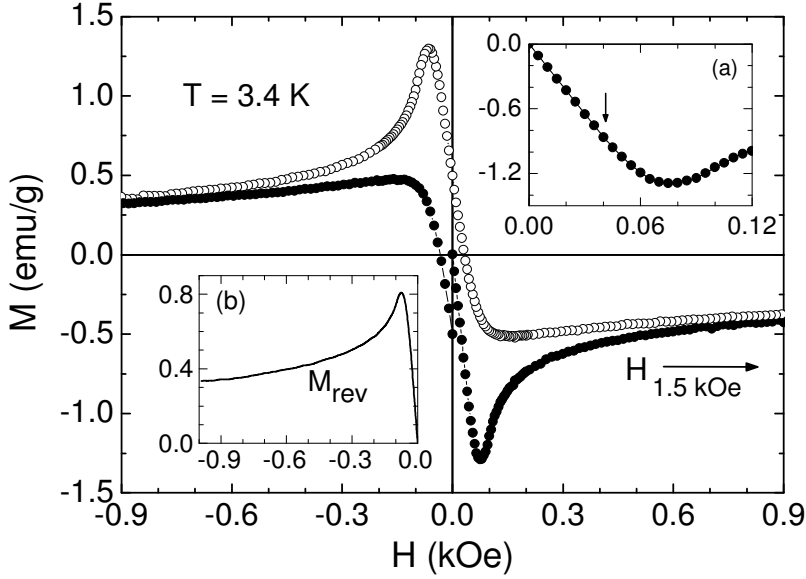

FIG. 6. Magnetization loop $M(H)$ obtained at 3.4 K for $H$ swept from zero to 1.5 kOe (closed circles), then to -1.5 kOe (open circles), and then back to zero (closed circles). The inset (a) shows the virgin curve of the $M(H)$ loop, which was used to determine the first penetration field, $H_{fp}$. The first deviation from the linear part of the virgin curve is marked by an arrow. The inset (b) shows the reversible magnetization, $M_{rev}$, calculated from the $M(H)$ half-loop (see text).

Macroscopic properties of superconductors in the mixed state depend on the internal structure of an individual vortex, which is determined by the GL parameter $\kappa = \lambda/\xi$. Both $\lambda$ and $\xi$ can be field dependent and thus the theoretical description of the macroscopic properties can become more complicated. For example, the London approach, which is a powerful tool in describing the thermodynamically reversible magnetization, has to be modified to fit the experimental data well, as shown for several high-$\kappa$ superconductors.[31] Hao and Clem proposed a model based on a variational method, which resulted in a precise description of the reversible magnetization and in determination of the thermodynamic critical field, $H_c$, and $\kappa$ as fitting parameters.[32] Recently, $M_{rev}(H)$ curves have been used to derive the GL parameters directly from the GL theory by applying the two-band model[33] and the two-order-parameter theory[34] with respect to $MgB_2$ and $YBa_2Cu_3O_7$, respectively.

General expression for $M_{rev}(H)$ curves for type-II superconductors has been proposed by Brandt for both triangular and square flux-line lattices.[35] This expression, which was achieved within GL theory, has been postulated to be valid for the entire range of applied fields $H_{c1} \leq H \leq H_{c2}$. Application of this and other models to analyze the $M_{rev}(H)$ data obtained for our $RbOs_2O_6$ and $KOs_2O_6$ single crystals are restricted due to the large demagnetizing factor of these octahedral-shaped samples. So we explain our results with caution being aware that using more developed models or more detailed analyses may result in misinterpretation of the experimental findings.[36] Additionally, as shown in many experiments, at low fields the irreversibility is dominated by geometrical barrier which renders the determination of the reversible part of the magnetization difficult. However, for fields larger than a few $H_{fp}$, where $H_{fp}$ is the first penetration field, the irreversibility is attributed mainly to the bulk pinning and the geometrical barrier effect can be neglected. $H_{fp}$ has been estimated to be about 82 and 95 Oe, for $RbOs_2O_6$ (this work) and $KOs_2O_6$ (Ref. 7) crystals, respectively. Thus, the reversible magnetization is considered only for $H > 200$ Oe $> 2H_{fp}$.

Magnetization loops similar to those shown in Fig. 6 have been used to derive the reversible magnetization curves $M_{rev}(H)$, defined as the average between ascending and descending branches of the magnetization loop. Figure 6 (inset) shows the $M_{rev}(H)$ curve at 3.4 K for the $RbOs_2O_6$ crystal, for the negative-field side of the $M(H)$ loop. A similar $M_{rev}(H)$ curve can be obtained for the positive-field side of the loop, if the half-loop instead of the virgin curve (shown in the Figure) is considered. This positive-field $M_{rev}(H)$ curve is taken for further analyses and plotted in a semi-log scale in Fig. 7, together with results obtained for $KOs_2O_6$ crystals. For $RbOs_2O_6$, a linear $M_{rev}(H)$ dependence is observed in a field range 0.2 kOe $< H <$ 1.5 kOe, while for $KOs_2O_6$, no linear dependence is found at any temperatures and fields studied. In the intermediate field range (London regime), the reversible magnetization, which is entirely defined by $h = H/H_{c2}$ and $\kappa = \lambda/\xi$, is expected to vary linearly with $\ln H$:[35]

$$4\pi M_{rev} = (-H_{c2}/4\kappa^2)\ln(0.358/h), \qquad (1)$$

where $M_{rev}$ is in emu/cm$^3$ and $H_{c2}$ in Oe. According to calculation of Brandt,[35,37] the basic requirement to observe a linear $M$ on $\ln H$ dependence is $H < 0.1H_{c2}$. At low temperatures ($T < 0.5T_c$), $H_{c2}$ for $RbOs_2O_6$ and $KOs_2O_6$ is higher than 30 and 100 kOe, respectively. Thus, the $M_{rev}(H)$ curves, which were obtained at fields up to 1.5 and 5 kOe for $RbOs_2O_6$ and $KOs_2O_6$, respectively, can be used to calculate the GL parameter $\kappa$ with formula (1). For $T = 3.4$ K and $H = 0.5$ kOe (see Fig. 7), a volume magnetization $M_{rev} = -3.47$ emu/cm$^3$ is obtained with a crystal density $\rho = 7.15$ g/cm$^3$. For $T = 3.4$ K, $H_{c2}$ is equal to 28.5 kOe (see Fig. 5) and, consequently, the GL parameter $\kappa = 22.2$ is calculated. For fields H $\geq$ 0.5 kOe, the correction for demagnetizing effects ($N = 0.5$), $H_{cor} = H - N4\pi M$, results in an increase of $H$ for less than 5 % and in a decrease of $\kappa$ for less than 1%, so we neglect this correction in most aspects of further consideration.

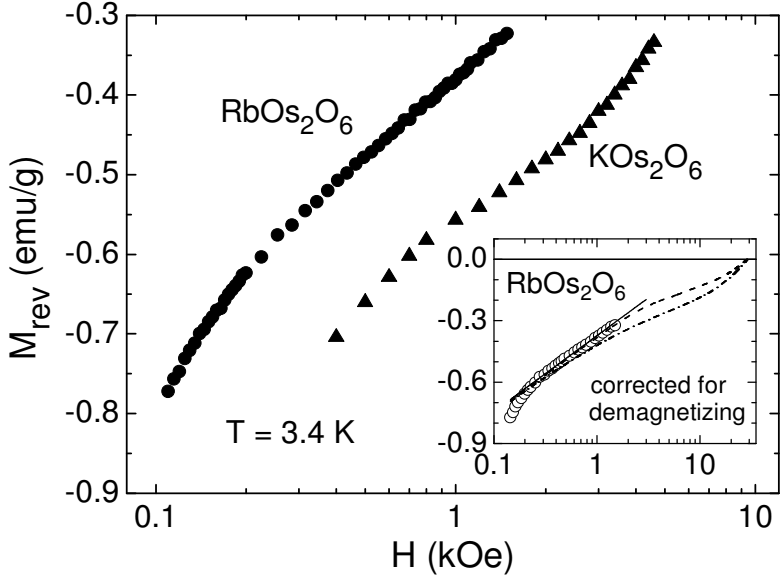

FIG. 7. Reversible magnetization as a function of an applied field, for $RbOs_2O_6$ and $KOs_2O_6$ single crystals. The data for $KOs_2O_6$ are shifted with +0.1 emu/g for convenience. The inset shows the experimental $M_{rev}(H)$ data (circles) corrected for demagnetizing effects (see text). The experimental data are compared to the results calculated with Eq. (1) for $\kappa(H)$ = const = 22.2 (solid line) and with Eq. (2) for the same $\kappa$ (dot-dashed line), and for $\kappa$ being field dependent (dashed line). The dashed line has been obtained for $\kappa$ changing gradually from 22.2 ($H \sim 0.2$ kOe) to 26.4 ($H \geq 13$ kOe).

The linear $M_{rev}(\ln H)$ dependence, which is observed for $RbOs_2O_6$ crystals, shows that $\kappa$ is field independent at low temperatures and fields from 0.25 to at least 2 kOe. Thus, first we assume that $\kappa(H)$ = const in the entire field range ($H \leq H_{c2}$) and calculate $M_{rev}(H)$ from the interpolation formula obtained by Brandt:[35]

$$4\pi M_{rev} = (-H_{c2}/4\kappa^2)\ln\{1+[(1-h)/h](0.357+2.890h-1.581h^2)\}. \qquad (2)$$

The calculated $M_{rev}(H)$ is shown in the inset of Fig. 7 revealing a rather poor fitting to the experimental results. For better fitting, $\kappa$ is required to be filed dependent and to change gradually from 22.2 at low fields to 26.4 at $H \geq 13$ kOe. Assuming that $\kappa$ is roughly proportional to $H$, we can obtain the extrapolated $\kappa(H_{c2}) \approx 31$ at 3.4 K, where $H_{c2} = 28.5$ kOe. This $\kappa$, which is expected to be temperature independent or weak-dependent at $T < 0.5T_c$, is very close to $\kappa(0) \approx 32$ derived from $H_{c2}(0)$, as described in the paragraph devoted to the GLAG theory. This may prove applicability of the standard theories and points to the classical behavior of the $RbOs_2O_6$ superconductor. Similar $\kappa(0) = 34$ was obtained for the polycrystalline compound from the specific heat measurements.[13]

Now, we can estimate the thermodynamic critical field $H_c(0) = H_{c2}(0)/\sqrt{2}\kappa \approx 1.3$ kOe, with the GL parameter $\kappa = 31$ and $H_{c2}(0) = 56$ kOe, which is a value obtained from the extrapolation of the $H_{c2}(T)$ experimental results (see Fig. 5). The penetration depth $\lambda$ is related to $H_c$ and $\kappa$ by $\sqrt{2}H_c = \kappa\Phi_0/2\pi\lambda^2$, with which $\lambda_o = \lambda(0) \approx 240$ nm is calculated. This result is in excellent agreement with the magnetization and muon-spin rotation ($\mu$SR) measurements, where $\lambda_o \approx 230$ nm was found for polycrystalline material.[10] Later $\mu$SR experiments showed that $\lambda_o$ ranged from 250 to 300 nm and was magnetic field almost independent at low temperatures.[11] Thus, using $\lambda_o \approx 240$ nm and $\kappa = 31$, which are the values obtained by us from the analysis of the $M_{rev}(H)$ curves, the GL coherence length is estimated to be $\xi_o = \xi(0) = \lambda_o/\kappa \approx 7.7$ nm. This is in excellent agreement with $\xi_o$ derived from the $H_{c2}(T)$ results and is only slightly higher than 7.4 nm obtained for a polycrystalline compound.[13,18]

Considering still the $M_{rev}(H)$ curves for RbOs$_2$O$_6$ crystals (see Fig. 7), we would like to remind that the logarithmic derivative d$M_{rev}$/dln$H$ is field independent in the filed range from 0.25 to about 2 kOe. This derivative is expected to be proportional to the superfluid density, which is inversely proportional to the penetration depth. Thus, the $M_{rev}(H)$ behavior can be well described in the London regime and the RbOs$_2$O$_6$ compound appears to be a conventional superconductor. Note, that for KOs$_2$O$_6$ crystals, no clear linear dependence of $M_{rev}$ vs ln$H$ can be found, so $\lambda$ seems to be field dependent and the $M_{rev}(H)$ theoretical curve may fit the experimental data only in a very limited field range by using a unique set of $\kappa$ values. This may emphasize the unusual superconducting properties of the KOs$_2$O$_6$ compound in the mixed sate including the exceptionally large upper critical field.[8]

Encouraged by the fact that RbOs$_2$O$_6$ reveals as a conventional BCS superconductor, we estimate the energy gap at 0 K, $\Delta_o$, by using the empirical relation first described by Toxen:[38] $[2T_c/H_c(0)][-dH_c(T)/dT]_{T_c} = 2\Delta_o/k_BT_c$, where d$H_c(T)$/d$T$ is taken at $T_c$. This relation is usually valid for the weak-coupling regime,[39] so we assume it can provide us with reasonable results for RbOs$_2$O$_6$. The d$H_c(T)$/d$T$ value at $T_c$ was derived from the relation $H_c(T) = H_{c2}(T)/\sqrt{2}\kappa$ with -d$H_{c2}(T)$/d$T$ = 11 kOe/K at $T_c$ = 6.13 K (see Fig. 5) and with $\kappa(T_c) = 23$ taken from Ref. 13. For these data, $2\Delta_o/k_BT_c \cong 3.2$ has been obtained, which is close to the weak-coupling BCS value of 3.53 and fits well into the range 3.1-3.6 found for the polycrystalline compound from the $\mu$SR measurements.[11] This result and all of the above-discussed relevant features confirm definitely that RbOs$_2$O$_6$ is a weak-coupling BCS superconductor.

Structural refinements of RbOs$_2$O$_6$ crystals reveal a large Rb isotropic displacement parameter at room temperature of 100x$U_{iso}$ = 3.57 and 4.31 Å$^2$, for Rb1 and Rb2, respectively (see Tab. II), when the refinements are done within the non-centrosymmetric space group $F\bar{4}3m$. These values are close to 100x$U_{iso}$ = 3.41 Å$^2$ reported for polycrystalline compound, when the structure was refined within the centrosymmetric space group $Fd\bar{3}m$.[9] The observed relatively large $U_{iso}$ exposes enhanced "rattling" of the Rb atom in the Os$_{12}$O$_{18}$ cage, if compare to the "rattling" of the Cs atom (100x$U_{iso}$ = 2.5 Å$^2$),[21] as predicted for the A'Os$_2$O$_6$ (A' = Cs, Rb, K) family from the band structure calculations.[40] These calculations postulate the largest displacement of the alkali atom for the $KOs_2O_6$ compound, where a substantial instability in the optic mod of the K atom leads to the extensive "rattling" behavior. This "rattling" is expected to create highly anharmonic low frequency phonons, which may influence the transport and thermodynamic properties significantly.[22,40,41] For single crystals, $100 \times U_{iso} \approx 6.3$ and $7.4$ Å² was obtained for the K atom, when the structure was refined in the space group $F\bar{4}3m$ (Ref. 7) and $Fd\bar{3}m$ (Ref. 21), respectively.

The large dissimilarity in the "rattling" intensity of the alkali metal atoms in the $\beta$-pyrochlores seems to be the main reason ("structural") for different superconducting properties observed for $A'Os_2O_6$ with $A'$ = Cs, Rb, and K. This has been indicated in the work by Hiroi et al.,[19] where systematic variations of the normal-state and superconducting properties have been considered over the Cs-Rb-K series. Figure 8 shows $T_c$ and $H_{c2}(0)$ versus $U_{iso}$ of the A' atoms, for polycrystalline $CsOs_2O_6$ and single crystals of $A'Os_2O_6$ (A' = Rb, K). A roughly linear increase of $T_c$ with $U_{iso}$ is observed across the Cs-Rb-K series. For $RbOs_2O_6$, the linear $T_c(U_{iso})$ dependence better applies to atoms which occupy the position

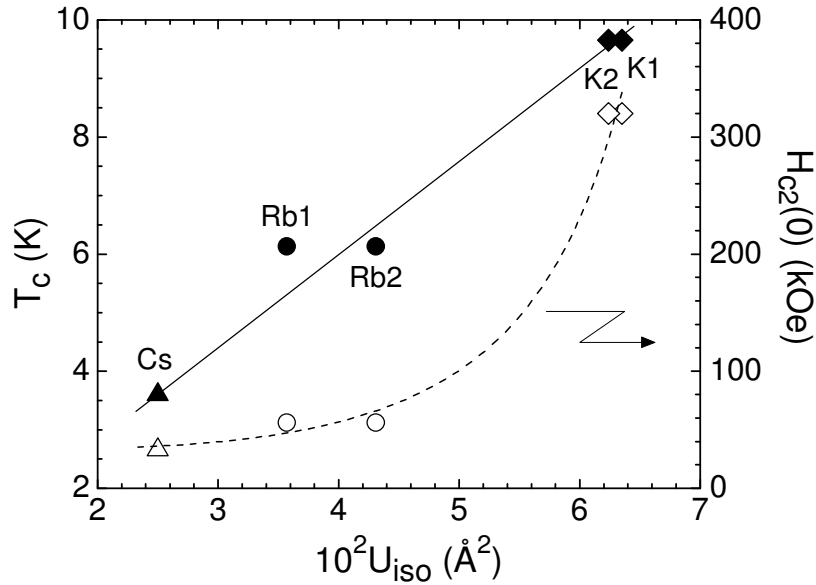

FIG. 8. Superconducting transition temperature (closed symbols) and an upper critical field at zero temperature (open symbols) versus $U_{iso}$ at 300 K of the A' atoms of $A'Os_2O_6$, for polycrystalline $CsOs_2O_6$ (triangles) and single crystals of $RbOs_2O_6$ (circles) and $KOs_2O_6$ (diamonds). Results for $CsOs_2O_6$ were taken from Ref. 21. Results for $KOs_2O_6$ come from our previous work.[7]

A'2, i.e., the atoms which vibrate with larger amplitude in more spacious channels of the Os-O network. On the other hand, $H_{c2}(0)$ increases with $U_{iso}$ much steeper revealing that no scaling between $T_c$ and $H_{c2}$ is here possible in the frame of the BCS theory. Since the crystal and electronic structures of $A'Os_2O_6$ (A' = Cs, Rb, K) are essentially the same, and since $U_{iso}$ is a measure of the instability of the alkali atom in the tetrahedral interstitial site, the observed monotonic relations $U_{iso}$-$T_c$ and $U_{iso}$-$H_{c2}$ suggest that the superconducting properties of these compounds depend more on the lattice dynamics than on the electronic structure. This is consistent with the results of the electronic structure investigations, where the different degree of anharmonicity of the alkali atom across the K-Rb-Cs series is proposed to be responsible for differences in the physical properties of these pyrochlores.[3,40] The lattice dynamics can be also responsible for the breaking of the inversion symmetry of the ideal β-pyrochlore structure to reduce the structural frustration predicted to appear for the tetrahedral sublattice of the alkali atoms.[3] Note, that the non-centrosymmetric structure has been proposed for $KOs_2O_6$ (Ref. 7) and is considered for $RbOs_2O_6$ in this work.

## V. CONCLUSION

In this paper we have examined the structural and superconducting properties of the high quality $RbOs_2O_6$ single crystals revealing that $RbOs_2O_6$ is a conventional superconductor, despite of a non-centrosymmetric structure refined at room temperature. The anisotropic character of $RbOs_2O_6$ is less pronounced than for $KOs_2O_6$, however additional reflections which violate the $Fd\bar{3}m$ symmetry have been certainly observed and the non-centrosymmetric $F\bar{4}3m$ space group has been chosen for interpretation of the crystallographic data. The high quality of $RbOs_2O_6$ crystals allowed us to refine anisotropic displacement parameters for all atoms. This refinement shows a strong motion ("rattling") of the Rb atoms in the framework of $OsO_6$ octahedra. The Rb atoms "rattle" in channels, which are anisotropic due to the breathing mode like behavior of the octahedral and tetrahedral networks. The lower displacement of the Rb atoms compared with K atoms in $A'Os_2O_6$ is in good agreement with band structure calculations where instability and motion anharmonicity of alkali metal atoms increase toward the Cs-Rb-K series of the β-pyrochlores.[40] The superconducting properties of the $RbOs_2O_6$ crystals in the mixed state are well described within the London approach revealing that in this compound the superconductivity is rather conventional. The superconducting gap $2\Delta_0/k_BT_c \cong 3.2$ has been estimated, indicative of the weak-coupling BCS-type pairing. The values of the coherence length, the GL parameter, and the lower and upper critical fields for single crystals and polycrystalline samples are similar, except of a clear difference observed for the initial slope of $H_{c2}(T)$. These results show that $RbOs_2O_6$ is slightly sensitive to small inhomogeneities and other imperfections, so $RbOs_2O_6$ seems to be a superconductor close to the dirty limit. For the $A'Os_2O_6$ family, both $T_c$ and $H_{c2}$ increase with $U_{iso}$ toward the Cs-Rb-K series supporting the belief by some that the superconducting properties of the β-pyrochlores depend more on the lattice dynamics than on the electronic structure which for all these compounds is very similar.

ACKNOWLEDGEMENTS

This work was supported by NCCR MaNEP and National Science Foundation. The SMART CCD measurements were performed at the Laboratory of Inorganic Chemistry, ETH Zürich. We acknowledge fruitful discussions with M. Wörle (Laboratory of Inorganic Chemistry, ETH Zürich).

**Note added in proof:**

Recently, the crystal structure of $KOs_2O_6$ and $RbOs_2O_6$ has been studied on our single crystals by Schoenes et al.,[42] using the micro-Raman spectroscopy. The results show that the structure is centrosymmetric in both compounds. Thus, the reflections which violate the centrosymmetric structure we observed could be due to the Renninger effect.[43] If this is true, no symmetry breaking is present and the structure of $KOs_2O_6$ and $RbOs_2O_6$ belongs to the centrosymmetric $Fd\bar{3}m$ space group. To clarify this issue, advanced studies of the crystal structure of these $\beta$-pyrochlores have been undertaken using synchrotron $X$-ray experiments.

TABLE I. Crystallographic and structure refinement parameters of β-pyrochlore $RbOs_2O_6$ at room temperature. The refinement was performed with Siemens SMART CCD diffractometer data (graphite-monochromated Mo $K\alpha$ radiation, $\lambda$ = 0.71073 Å). The edge length was 0.08 mm. The calculated material density $\rho$ = 7.193 g/cm$^3$ and the calculated absorption coefficient $\mu$ = 58.200 mm$^{-1}$.

| Space group (Crystal system) | $F\bar{4}3m$ ( cubic ) |
|---|---|
| Unit cell $a$ (Å); Z | 10.1242(12); 8 |
| Volume (Å$^3$) | 1037.7(2) |
| θ range (deg) | 3.49 - 33.79 |
| $h_{min}$; $k_{min}$; $l_{min}$ | -15; -15; -15 |
| $h_{max}$; $k_{max}$; $l_{max}$ | 15; 15; 15 |
| Extinction coefficient | 0.00321(13) |
| Flack parameter | 0.61(12) |
| Total Reflections measured | 4053 |
| Symmetry equivalents reflection; Variables | 249; 16 |
| $R_{int}$; $R_{sig}$ | 0.0571; 0.0234 |
| $R[F^2 > 2\sigma(F^2)]$; $wR(F^2)$ | 0.0194; 0.0371 |
| S (observed data); $(\Delta/\sigma)_{max}$ | 1.445; 0.000 |
| Largest $\Delta\rho$ (x,y,z),max; min (e/Å$^3$) | 2.997; -1.035 |

TABLE II. Atomic coordinates, anisotropic ($U_{ij}$) and isotropic ($U_{iso}$) displacement parameters (in Å$^2$) for $RbOs_2O_6$ and, to compare, $U_{iso}$ for $KOs_2O_6$ (Ref. 7). The results were obtained at room temperature with the non-centrosymmetric space group $F\bar{4}3m$.

| $RbOs_2O_6$ | Os | Rb1 | Rb2 | O1 | O2 |
|---|---|---|---|---|---|
| site | 16e | 4c | 4b | 24f | 24g |
| X | 0.87579(5) | ¼ | ½ | 0.1899(12) | 0.557(2) |
| Y | X | ¼ | ½ | 0 | ¼ |
| Z | X | ¼ | ½ | 0 | ¼ |
| $U_{11}$ | 0.0093(2) | 0.0357(14) | 0.0431(16) | 0.001(4) | 0.036(7) |
| $U_{22}$ | 0.0093(2) | 0.0357(14) | 0.0431(16) | 0.013(3) | 0.019(4) |
| $U_{33}$ | 0.0093(2) | 0.0357(14) | 0.0431(16) | 0.013(3) | 0.019(4) |
| $U_{23}$ | 0.00066(6) | 0 | 0 | 0.004(4) | 0.014(4) |
| $U_{13}$ | -0.00066(6) | 0 | 0 | 0 | 0 |
| $U_{12}$ | 0.00066(6) | 0 | 0 | 0 | 0 |
| $U_{iso}$ | 0.0093(2) | 0.0357(14) | 0.0431(16) | 0.0087(17) | 0.025(2) |
| **$KOs_2O_6$** | **Os** | **K1** | **K2** | **O1** | **O2** |
| $U_{iso}$ | 0.0005(1) | 0.063(5) | 0.062(4) | 0.005(1) | 0.005(1) |

TABLE III. Selected interatomic distances (Å) and bond angles (°) for RbOs$_2$O$_6$ and KOs$_2$O$_6$ (Ref. 7) obtained within the non-centrosymmetric space group $F\bar{4}3m$. The length difference, $\Delta$, caused by lowering of the symmetry from $Fd\bar{3}m$ (ideal β-pyrochlore structure) to $F\bar{4}3m$, is defined as: $\Delta$ = (maximum length) - (minimum length). Numbers in square brackets indicate bond multiplicities.

| β-pyrochlore | | RbOs$_2$O$_6$ | Δ (Å) | KOs$_2$O$_6$ | Δ (Å) |
|---|---|---|---|---|---|
| **osmium environment** | | | | | |
| Os – O1 | [3] | 1.899(4) | 0.026 | 1.920(5) | 0.014 |
| Os – O2 | [3] | 1.925(7) | | 1.906(4) | |
| O1 – Os – O1 | | 91.5(5) | | 94.2(5) | |
| O1 – Os – O2 | | 88.3(2) | | 87.8(3) | |
| O2 – Os – O2 | | 91.9(8) | | 90.1(5) | |
| O1 – Os – O2 | | 179.7(9) | | 177.0(6) | |
| Os – O1 – Os | | 139.0(7) | | 137.7(3) | |
| Os – O2 – Os | | 138.4(11) | | 138.9(6) | |
| **tetrahedral network** | | | | | |
| Os – Os | [4] | 3.5568(8) | 0.0453 | 3.548(4) | 0.044 |
| Os – Os | [4] | 3.6021(8) | | 3.592(5) | |
| **alkali metal (A') environment** | | | | | |
| A'1 – O2 | [8] | 3.1081(4) | 0.0314 | 3.141(13) | 0.082 |
| A'2 – O1 | [8] | 3.1395(4) | | 3.059(13) | |
| A'1 – A'2 | [4] | 4.3839(3) | | 4.3720(4) | |
| **octahedral network: OsO$_6$ octahedra** | | | | | |
| O1 – O2 | [6] | 2.666(5) | | 2.653(4) | |
| **octahedral network: O$_6$ octahedra** | | | | | |
| O1 – O1 | [8] | 2.719(12) | 0.041 | 2.819(5) | 0.124 |
| O2 – O2 | [8] | 2.76(2) | | 2.695(4) | |